\RequirePackage{fix-cm}
\documentclass[twocolumn,epjc3]{svjour3}  
\smartqed  
\RequirePackage{graphicx}
\RequirePackage[numbers,sort&compress]{natbib}
\RequirePackage{amsmath}
\RequirePackage{color}
\newcommand{\ba}{\begin{eqnarray}}
\newcommand{\ea}{\end{eqnarray}}
\newcommand{\bea}{\begin{eqnarray}}
\newcommand{\eea}{\end{eqnarray}}
\newcommand{\nn}{\nonumber}

\newcommand{\be}{\begin{equation}}
\newcommand{\ee}{\end{equation}}

\newcommand{\eq}[1]{(\ref{#1})}
\journalname{Eur. Phys. J. C}
\begin{document}

\title{GW170817 and GW190425 as Hybrid Stars of Dark and Nuclear Matter
}


\author{Kilar Zhang\thanksref{e1,addr1,addr2,addr3}
        \and
        Guo-Zhang Huang\thanksref{e2,addr1} 
        Jie-Shiun Tsao\thanksref{e3,addr1} 
        \and
        Feng-Li Lin\thanksref{e4,addr1,addr4} 
}

\thankstext{e1}{kilar.zhang@gmail.com}
\thankstext{e2}{60641027s@ntnu.edu.tw}
\thankstext{e3}{tsaojieshiun@gmail.com}
\thankstext{e4}{linfl@gapps.ntnu.edu.tw Corresponding Author}


\institute{Department of Physics, National Taiwan Normal University, Taipei 11677, Taiwan \label{addr1}
           \and
            Department of Physics, Shanghai University, Shanghai 200444, Mainland China\label{addr2}
            \and
           Shanghai Key Lab for Astrophysics, Shanghai 200234, Mainland China\label{addr3}
            \and
           Center of Astronomy and Gravitation, National Taiwan Normal University, Taipei 11677, Taiwan\label{addr4}
}

\date{Received: date / Accepted: date}

\maketitle

\begin{abstract}
We propose three scenarios for compact hybrid stars composed of nuclear and dark matter.  These hybrid stars could provide alternative interpretations to the LIGO/Virgo events GW170817 and GW190425. To demonstrate our proposal, we solve the Tolman-Oppenheimer-Volkoff configurations of hybrid stars by using the SLy4, APR4, and SKb equations of state (EoS) for nuclear matter, and an EoS for a bosonic self interacting dark matter (SIDM) proposed by Colpi et al \cite{Colpi:1986ye}. We then obtain their mass-radius and tidal Love number (TLN)-mass relations, and further examine the possible saddle instability of these compact objects by the generalized Bardeen-Thorne-Meltzer (BTM) criteria. Our results show that the hybrid star scenarios are able to explain GW170817 and GW190425. Some hybrid stars can have compact neutron or mixed cores around 10 km while possessing thick dark matter shells, thus they can be more massive than the maximum mass of the typical neutron stars but are electromagnetically detected with about the same size of neutron stars. Reversely, we also infer the dark matter model from the parameter estimation of GW190425. Our proposed hybrid stars can be further tested by the coming LIGO/Virgo O3 events.
\end{abstract}

\section{Introduction}
\label{intro}
Dark matter, though constituting 85\% of the matter content of the Universe, reveals little evidence via direct search of the past three decades \citep{Klasen:2015uma,Akerib:2016vxi,Cui:2017nnn,Aprile:2017iyp}. However, it is hard to detect the dark matter directly by the conventional electromagnetic means due to its rare interaction with baryonic matter. On the other hand, everything gravitates. If dark matter can form compact stars, the gravitational wave (GW) emitted from the associated compact binary coalescence (CBC) can then be detected. By the observed data, we can infer the equation of state (EoS) \citep{Malik:2018zcf,LIGOScientific:2018cki,De:2018uhw,Tews:2018iwm,Radice:2017lry, HernandezVivanco:2020cyp} and the corresponding  microscopic theory of dark matter \citep{Kouvaris:2015rea,Maselli:2017vfi,Sennett:2017etc}. This can be thought as an alternative direct search through the relation between gravitational astronomy, microscopic and macroscopic physics of dark matter. The most popular model for dark matter is the WIMP (weakly interacting massive particles) \citep{Jungman:1995df}. Despite that, the WIMP cannot well explain some astrophysical properties of the dark matter halo, such as the smooth core profile or the missing satellites. It motivates to introduce SIDM (the self-interacting dark matter) to resolve these issues \citep{Spergel:1999mh,Rocha:2012jg,Peter:2012jh,Kaplinghat:2015aga}.   Moreover, it has been shown \citep{Colpi:1986ye,Schunck:2003kk,Eby:2015hsq,Deliyergiyev:2019vti} that some bosonic models of SIDM can yield compact stars of a few solar masses, the so-called dark stars. 

\begin{figure}[htpb]
\resizebox{\hsize}{!}{\includegraphics{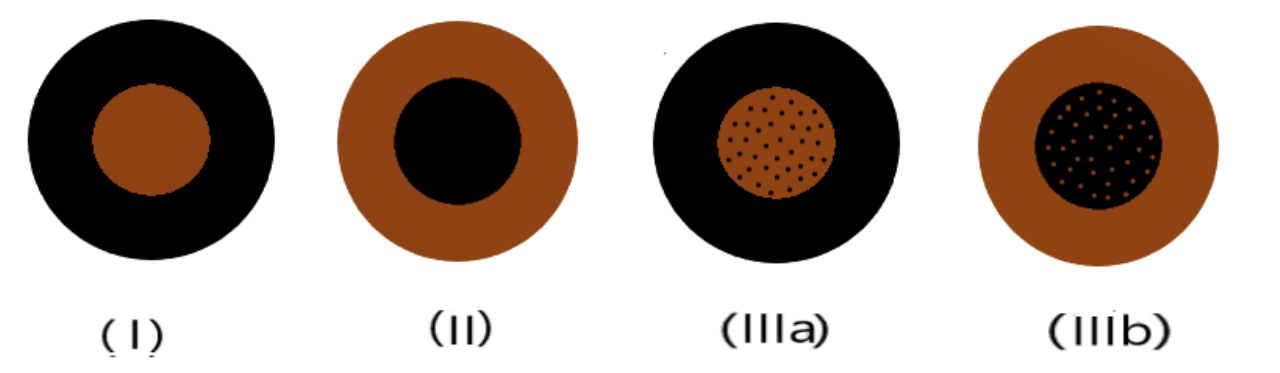}}
  \caption{Three scenarios of hybrid stars. Black color denotes dark matter and brown color denotes nuclear matter. For scenario I the star consists of a pure nuclear matter core and a pure dark matter shell, and for scenario II it consists of a pure dark matter core and a pure nuclear matter shell. For scenario III, we have a mixed core and either a pure dark matter shell (IIIa) or a pure nuclear matter one (IIIb). They can form the systems of binary hybrid stars (BHS). }\label{stars}
\end{figure}

Given the possibility of compact dark stars, one can speculate the existence of compact hybrid stars composed of both dark and nuclear matter. This is the analogue to the dark halos made of dark and baryonic matter \citep{Battaglia:2005rj} but in a much smaller scale. There are three scenarios of compact hybrid stars as shown in Fig. \ref{stars}, which depend on how dark and nuclear matter interact, and also on the accretion mechanism. Scenario I is to have the stars with neutron core and dark matter shell, and Scenario II is for the stars with dark matter core and neutron shell. For both scenarios we assume there is either interaction between dark and nuclear matter, or spontaneous symmetry breaking to form the domain wall separating the core and the shell \footnote{A possible scenario for the formation of domain wall between nuclear matter and dark matter is for them to belong to two different symmetric phases of the same underlying theory, such as some kind of grand unified model.  We admitted that the scenario I and II remain hypothetical without a concrete model realization. In \citep{Gresham:2018rqo} a domain wall can form between fermionic dark matter and nuclear matter with a fine-tuned repulsive interaction but a negligible cross section $10^{-47} \textrm{cm}^2$, see the rightmost plot of Fig. 9 of \citep{Gresham:2018rqo}. This can be seen as a dynamical alternative to the above phase separation scenario.}. Otherwise, it will lead to  scenario III, in which the dark and nuclear matter mix inside the core, but with only one component in the shell. Some models for hybrid stars of scenario IIIa and IIIb have been proposed and studied in \citep{Nelson:2018xtr} and \citep{Ellis:2018bkr}, respectively. These hybrid stars can be seen as the cousins of neutron stars, with the new parameter $r_W$ characterizing the radius of  the inner core. Here, we simply assume the existence of these hybrid stars and leave their formation mechanism for future studies. \footnote{The capture rate of dark matter by neutron star is too small to form sizable share in the hybrid stars due to the negligible interaction between dark matter and baryons \citep{Kouvaris:2010jy,McDermott:2011jp}. However, there are proposals to dramatically increase the cross-section by forming the nuggets \citep{Gresham:2017zqi, Gresham:2017cvl,Gresham:2018anj,Coskuner:2018are} so that the capture rate could be accelerated. Another possibility for quickly accumulating dark matters is through the Bondi accretion \citep{Edgar:2004mk}.   }

In contrast to neutron stars, it is hard to directly observe dark stars or hybrid stars by electromagnetic signals since the dark matter couples very weakly to the baryonic matter. Despite that, some key properties such as mass, radius, and tidal deformability of the compact stars depend sensitively on the details of the constituent matter, including the proportions of the nuclear and dark matter, and their associated equations of state. Therefore, we may be able to identify the dark or hybrid stars by measuring these key properties, which will be encoded in the GW emitted from the coalescence of binary hybrid stars (BHS).
In such a way, we can test the above three scenarios of BHS via GW events discovered by LIGO/Virgo. Currently, there are two observed events usually identified as binary neutron stars (BNS), namely, GW170817 \citep{TheLIGOScientific:2017qsa,Abbott:2018wiz} and GW190425 \citep{Abbott:2020uma}. The key feature of both events is their low tidal Love numbers (TLNs). Moreover, the total mass and the associated component masses of GW190425 seem larger than the ones expected for the neutron stars, in comparison to GW170817. Besides, in the event GW190814 \citep{Abbott:2020khf}, one of its component stars is inferred to have a mass of 2.6 $M_{\odot}$. This is marginally heavier than the typical maximum mass of neutron stars. 
Thus, this component star can hardly be considered as an ordinary neutron star, instead it could be either a dark or a hybrid star. Therefore, due to the lack of electromagnetic signals for the above GW events, they all can be the candidates of dark or hybrid stars. Our study in this work is trying to provide a framework to examine such possibilities in details. We hope that future observations will scrutinize these scenarios.  Reversely, we can constrain the dark matter model by analyzing the astrophysical properties of the hybrid stars from the GW data.\footnote{Note added: Following the main idea of this paper, in our subsequent work \citep{Zhang:2020dfi} we extend to other dark matter models and try to interpret GW190814 \citep{Abbott:2020khf} as the hybrid stars proposed here. In particular, some $M$-$R$ and $\Lambda$-$M$ relations presented in this paper and discussion of BTM criteria are adopted in \citep{Zhang:2020dfi}.}

 The organization of the paper is as follows: In Section 2 we introduce the EoS for the models used for dark and nuclear matters; Section 3 shows TOV configurations and how to calculate the tidal Love number; Section 4 illustrates $M$-$R$ and $\Lambda$-$M$ relations from our numerical results; 
 In Section 5 we fit GW170817 and GW 190425 using our models;  Section 6 describes the parameter estimation for the EoS of dark matter; Section 7 is the conclusion. In the Appendix, we summarize the (reverse) BTM Stability Criteria.

\section{Model of equation of state for dark and nuclear matter} 
In this paper, we show that both GW170817 and GW190425 can be easily explained by a simple toy model of hybrid stars based on the above three scenarios. We do not aim to pin down the models for nuclear and dark matter, but to demonstrate the viability of the hybrid star scenarios.  There are many bosonic dark matter model candidates, some of which predict light dark matter particles such as Axion \citep{Kuster:2008zz}. However, the ultralight dark matter is difficult to form compact stars \citep{Eby:2019ntd}, instead it may form dark matter clouds around compact objects \citep{Hannuksela:2018izj}.

For our purpose of searching for compact stars of a few solar masses, it is more natural to choose heavier dark matter with mass around MeV to GeV. It had been shown that the WIMP may not be able to yield smooth dark halo profiles. This leads to the proposal of introducing SIDM to fix the issue \citep{Spergel:1999mh,Rocha:2012jg,Peter:2012jh,Kaplinghat:2015aga}. Thus, when considering a model of hybrid stars, we adopt a simple SIDM model, that is a bosonic complex scalar field $\phi$ with potential $m^2 |\phi|^2+{\lambda\over 4} |\phi|^4$, where $m$ is the dark matter mass and $\lambda$ is the coupling constant. Moreover, we consider this model in an extreme regime with $\lambda M^2_{planck} /m^2 \gg 1$, where $M_{planck}$ is the Planckian mass scale. In this regime, this SIDM model can be equivalently described by a hydrodynamic perfect fluid with EoS given by \citep{Colpi:1986ye}
\be\label{DEoS}
\rho/\rho_{\odot}=3 \; (p/p_{\odot})+{\cal B}\; (p/p_{\odot})^{1/2}\,,
\ee
where ${\cal B}\sim {0.08 \over \sqrt{\lambda}}({m\over \textrm{GeV}})^2$ is a free  parameter\footnote{ 
This EoS is slightly below the ``sound barrier", which means the derived sound speed is always below the conformal limit, i.e., $1/\sqrt{3}$ of the light speed, and it takes the same form as the conjectured EoS of quark matter in the deep core of neutron star \citep{Hoyos:2016zke,Annala:2017tqz}. The coincidence is due to the $\phi^4$ potential and the nearly massless nature of quarks in the high chemical potential limit. If one consider higher $\phi^n$ with $n>4$, the resultant EoS can break sound barrier \citep{Zhang:2020dfi}. }. In this work, we adopt the following astrophysical units
\be\nn
r_{\odot}=G_N M_{\odot}/c^2,\quad \rho_{\odot}=M_{\odot}/r_{\odot}^3, \quad p_{\odot}= c^2 \rho_{\odot}
\ee
which are the (half) Schwarzschild radius of the Sun, the corresponding energy density and pressure, respectively. Note that $M_{\odot}$ denotes the solar mass, and $G_N$ the Newton constant, which is related to the Planckian mass by $G_N=1/M^2_{planck}$.

For this SIDM model to explain the smooth density profile of dark halos, one needs to impose the constraint on the cross-section of self-scattering, which is translated into a tiny window for $\lambda$ \citep{Kaplinghat:2015aga}:
\be\label{SIDMw}
30 ({m\over \textrm{GeV}})^{3/2} < \lambda < 90 ({m\over \textrm{GeV}})^{3/2}.
\ee
Therefore, if we can pin down the parameter ${\cal B}$ from the GW events, we can almost determine the parameters of this SIDM model.  As a first step, in this work we only consider the bosonic SIDM.  There are also fermionic SIDM \citep{Kouvaris:2015rea,Maselli:2017vfi,Gresham:2018rqo}, which we leave for the future work.

For the nuclear matter, there are also many  candidate models with different EoS. One type is the phenomenological model like the SLy4 \citep{Douchin:2001sv}, APR4 \citep{Akmal:1998cf} and SKb \citep{Gulminelli:2015csa} \footnote{Available from https://compose.obspm.fr} used extensively in gravitational wave data analysis \citep{TheLIGOScientific:2017qsa,Abbott:2018wiz,Abbott:2020uma}.  The other type is derived theoretically, like the one in \citep{Hirayama:2019vod} from a well-motivated holographic quantum chromodynamics model, i.e., Sakai-Sugimoto model \citep{Sakai:2004cn,Sakai:2005yt,Hata:2007mb}.  Since the phenomenological model is more generally accepted, we choose three representative phenomenological models for the discussion.
  
\section{Tolman-Oppenheimer-Volkoff  configuration and tidal Love number} \label{section 3}
  The GW of CBC encodes the component masses $M_{1,2}$, and also the TLNs $\Lambda_{1,2}$ in the following combined quantity
\be\label{tlambda}
\tilde{\Lambda}={16\over 13}{(M_1+12 M_2) M_1^4 \Lambda_1+(M_2+12 M_1) M_2^4 \Lambda_2 \over (M_1+M_2)^5}.
\ee 
Note that $\tilde{\Lambda}=(\Lambda_1+\Lambda_2)/2$ for $M_1=M_2$. For each hybrid star scenario, we have two model parameters $\cal B$ and $r_W$. We shall connect the model parameters  to the inferred quantities from the observation data by the mass-radius and TLN-mass relations. 

Given a set of $({\cal B}, r_W)$ we first obtain the mass-radius relation by solving the Tolman-Oppenheimer-Volkoff (TOV) equations \citep{Tolman:1939jz, Oppenheimer:1939ne}  for multi-component cases  \citep{Mukhopadhyay:2016dsg,Rezaei:2018cuk}  using units $G=c=1$:
\be\label{TOV-comp}
p'_I=-(\rho_I + p_I) \phi',\quad  m'_I =4\pi r^2 \rho_I, \quad \phi' ={m +4\pi r^3 p \over r(r-2m)},
\ee
where  $' :={d\over dr}$, $I=D$ or $N$, the mass inside radius $r$ is $m(r)=\sum_I m_I$,  pressure $p=\sum_I p_I$,  energy density $\rho=\sum_I \rho_I$ by summing the contributions from both dark matter ($I=D$) and nuclear matter ($I=N$), and the Newton potential $\phi:={1\over 2} \ln(-g_{tt})$ with $g_{tt}$ the $tt$-component of the metric. The size $R$ of the star is determined by $p (r=R)=0$, and the mass of the star is given by $m(R)$.
For the first scenario, we set $p_D=\rho_D=0$ and use EoS SLy4 to solve TOV equations for $r\le r_W$. For  $r\ge r_W$ we set $p_N=\rho_N=0$ and set initial value of $p_D$ at $r_W$ equal to $p_N(r_W)$, then use \eq{DEoS} to solve the TOV equations until $r=R$. For scenario II, we do the same thing by swapping the roles of dark and nuclear matters. For the third scenario, we  tune the initial values at $r=0$ for both $p_D$ and $p_N$ and use both \eq{DEoS} and EoS SLy4 to solve the TOV. In this case, $r_W$ is determined by the first vanishing $p_I$, then we solve the TOV equations for $r>r_W$ until $r=R$ for the remaining non-vanishing $p_I$ component. 

   Naively, one can sum \eq{TOV-comp} over the components to get a set of single-fluid TOV equations.  However, the EoS for each component is usually not in a linear form, it is then impossible to form an EoS of multi-fluid, namely $\rho=\sum_I \rho_I\ne \rho(p)$. This requires to impose an initial condition for each component when solving the TOV, and leads to the subtlety discussed right below when considering the stability of the hybrid stars of the third scenario based on the generalization of the Bardeen-Thorne-Meltzer (BTM) criteria \citep{1966ApJ...145..505B}. 

   A key difference between the first two scenarios and the third one is the stability issue. For the first two, we have just one initial value parameter for solving TOV, i.e., either $p_D(0)$ or $p_N(0)$, but have both for the third one, thus the stability issue of the latter is more tricky due to the possible saddle instability. 

Another issue is regarding the determination of $r_W$. It seems that $r_W$ is a free parameter in scenario I and II, but can be determined automatically in scenario III. This, however, is not true. For scenarios I and II to have a domain wall separating the dark matter and baryonic phases, there should have model interactions between these two components, which will then build up the chemical equilibrium between them. This implies that some mixed phase rule should be applied to determine $r_W$. Therefore, even in scenarios I and II, $r_W$ is not a free parameter. Despite that, it needs to specify the model interaction to determine $r_W$ unambiguously, such as the example in \citep{Gresham:2018rqo}. In this work we will not consider scenario I and II for any particular model interactions, instead we choose $r_W$ as a parameter to characterize the model interactions between dark matter and baryons.

  After having solved the stable TOV configurations, we solve the linear perturbation around them to extract the TLN, denoted by $\Lambda$ and defined by
\be
Q_{ab}=- M^5  \Lambda \; {\cal E}_{ab}\,,
\ee  
where $M$ is the mass of the star, $Q_{ab}$ is the induced quadruple moment, and ${\cal E}_{ab}$ is the external gravitational tidal field strength. To consider the tidal deformability for the hybrid stars of scenario III, we need to generalize the derivation of \citep{Hinderer:2007mb,Postnikov:2010yn} for the single fluid to the multi-fluid cases. We again need to solve the following equation for $y(r):=rH'(r)/H(r)$ with $H(r)$ the linear perturbation of $g_{tt}$ around a TOV configuration:
\be\label{TLN-y}
ry'+y^2+P(r)y+r^2 Q(r)=0\,,
\ee
with the boundary condition $y(0)=2$ and 
{
\bea
P&=&(1+4\pi r^2(p-\rho))/(1-2m/r), \qquad  \\
Q&=&4\pi (5\rho+9p+\sum_I \frac{\rho_I+p_I}{dp_I/d\rho_I}-\frac{6}{4\pi r^2})/(1-2m/r) \nonumber
\\&\qquad&-4\phi'^2. \qquad \label{multiQ}
\eea
}
The above equations are rigorously derived from Einstein equation, and the main difference from the single-fluid case is encoded in the $\sum_I \frac{\rho_I+p_I}{dp_I/d\rho_I}$ term of \eq{multiQ}.

Moreover, when we consider a hybrid star with two separated phases like the first two scenarios, the TOV and tidal equations must be solved separately for each phase, and then match the results by junction conditions on the domain wall. The pressure $p$ is continuous, but the energy density $\rho$ is not, so that  $y$  encounters a jump around the domain wall. The required junction conditions are sketched below, and for more details, see \citep{Postnikov:2010yn}.

Suppose the pressure reads $p_W$ on the domain wall located at $r=r_W$. The sound speed near a density discontinuity is
\begin{equation}
	\label{eqdendis}
	\frac{d\rho}{dp}=\frac{1}{c_s^2}=\left. \frac{d\rho}{dp}
	\right|_{p \neq p_W}+\Delta \rho_p \, \delta(p-p_W),
	\end{equation} 
where $\Delta \rho_p=\rho(p_W+0)-\rho(p_W-0)$ is the energy density jump across $p_W$. Yet since $p$ decreases as $r$ increases, equivalently $\Delta \rho_p=-\left(\rho(r_W+0)-\rho(r_W-0)\right)\equiv -\Delta\rho$.

	When integrating \eq{TLN-y} near the domain wall at $r=r_W$, most of the terms give zero, and only the terms proportional to the $\delta$-function can contribute. Therefore, this then results in
	\begin{equation}
ry^\prime(r)\big|_{r = r_W}+
r^24\pi e^{\lambda(r)} 
\left({\rho(r)+p(r)}\right)\frac{d\rho}{dp}\big|_{r = r_W}=0\,.
\end{equation}
Since  $\frac{d\rho}{dp}=\frac{d\rho}{dr}\frac{1}{dp/dr}$, where ${dp \over dr}$ can be read off from the first TOV equation \eqref{TOV-comp}, and 
$ \frac{d\rho}{dr}|_{r = r_W}=\Delta\rho \, \delta(r-r_W)$,
we obtain that \footnote{There are typos in the counterpart of \eqref{juntion} in \citep{Postnikov:2010yn} .}
\begin{equation}\label{juntion}
	\Delta y=\frac{\Delta \rho}{p+m(r_W)/(4 \pi r_W^3)}\,,
\end{equation}
where $\Delta y\equiv y(r_W+0)-y(r_W-0)$.
As a result, we can solve the TOV and tidal deformation equations with the above junction conditions for scenarios I and II.

Once \eq{TLN-y} is solved, the TLN $\Lambda$ can be obtained through an algebraic expression of $y_R\equiv y(R)$ and the ``compactness" $C=M/R$ given by \citep{Hinderer:2007mb,Postnikov:2010yn}
\begin{eqnarray}
&& \Lambda = \frac{16}{15}\left(1-2C\right)^2
\left[2+2C\left(y_R-1\right)-y_R\right]\times   \nonumber  \\
&&\bigg\{2C\left(6-3 y_R+3 C(5y_R-8)\right)\\
&& ~ ~+4C^3\left[13-11y_R+C(3 y_R-2)+2
C^2(1+y_R)\right] \nonumber\\
&& ~ ~
+3(1-2C)^2\left[2-y_R+2C(y_R-1)\right]\log\left(1-2C\right)\bigg\}^{-1}.\nonumber
\end{eqnarray}

\section{$M$-$R$ and $\Lambda$-$M$ relations} \label{section 4}         
\begin{figure}[htpb]
\resizebox{\hsize}{!}{\includegraphics{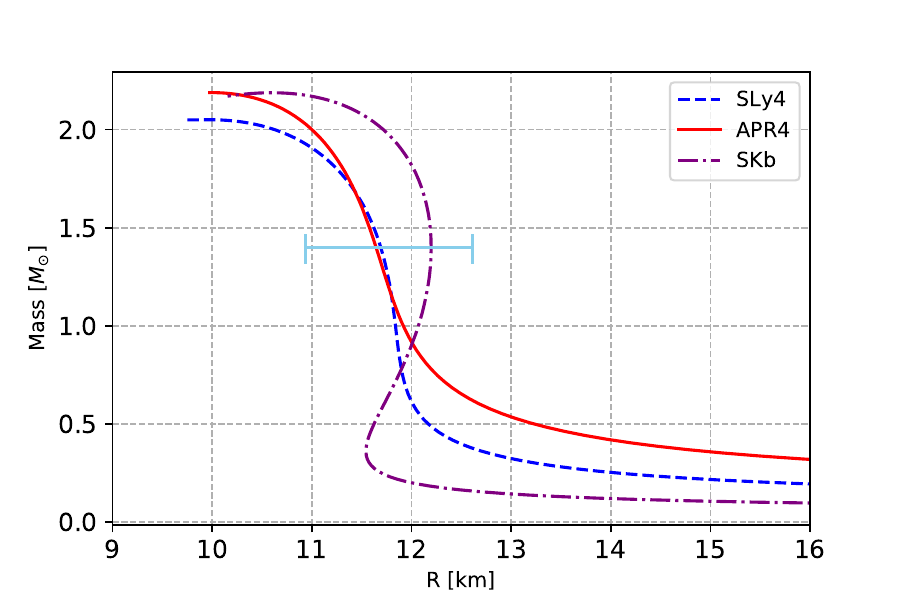}}
  \caption{Mass-Radius relations for three representative neutron EoS: SLy4, APR4 and SKb. The aqua line indicates the multimessenger constraints from the results of  \citep{Dietrich:2020efo} on the radii of neutron stars of 1.4 $M_{\odot}$. }\label{3EOS}
\end{figure}

Based on the above, we evaluate the $M$-$R$ and $\Lambda$-$M$ relations for the three hybrid star scenarios with dark matter EoS given by \eq{DEoS} and nuclear matter EoS given by SLy4, APR4 and SKb, respectively. In Fig. \ref{3EOS} we first show the $M$-$R$ relations of the three chosen neutron EoS, where we see they all yield the neutron stars of maximal mass slightly above 2$M_{\odot}$, and satisfy the multimessenger constraints given in \citep{Dietrich:2020efo}, i.e., the radius of a 1.4 $M_{\odot}$ neutron star should be 
$11.75^{+0.86}_{-0.81} \mbox{ km}$ at $90\%$ confidence level. These constraints are obtained assuming pure neutron stars. There are more possibilities when introducing the contribution of dark matter as considered in this work. 

\begin{figure*}
\sidecaption
\includegraphics[width=11cm]{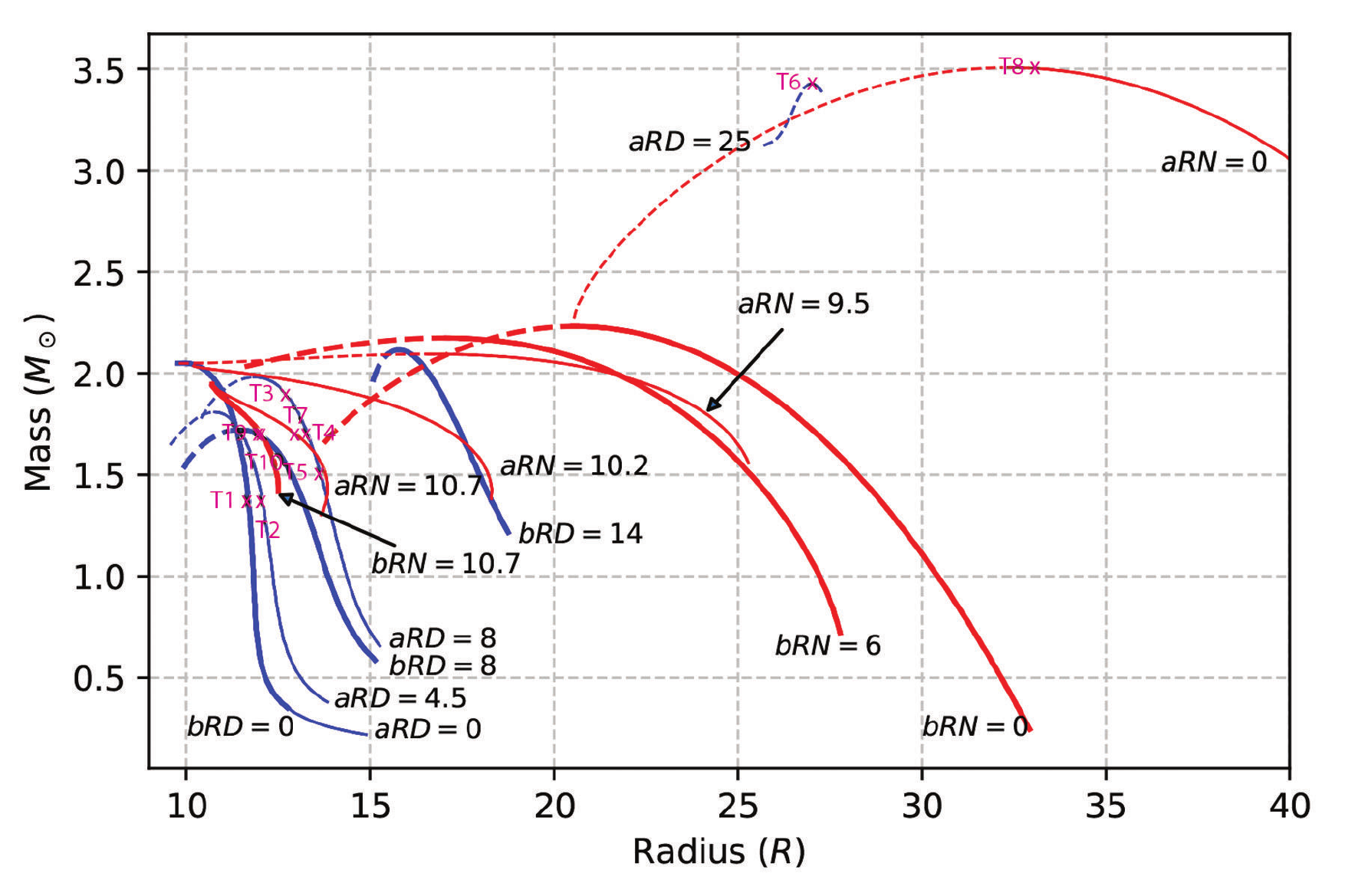}
\caption{Mass-Radius relations for the hybrid stars of type (I) and (II) in Fig. \ref{stars}, which are made of nuclear matter of SLy4 EoS and dark matter of EoS \eq{DEoS} with ${\cal B}=0.035$ or $0.055$. With ${\cal B}=0.035$ these relations are labelled  by {\bf aRN}$=r_W$ (red) for the first scenario, by {\bf aRD}$=r_W$ (blue) for the second. Similarly, with ${\cal B}=0.055$ they are labelled by {\bf bRN}$=r_W$ (red) and {\bf bRD}$=r_W$ (blue)). For example, {\bf aRD}=8 means the radius of dark core is $8 \textrm{km}$, with ${\cal B}=0.035$. The unstable configurations are indicated by the parts of dashed lines.  Note that $r_W$ is a parameter to characterize the model interaction between dark matter and baryons.
The crosses labeled by $T_n$ indicate special stars mentioned later in Table \ref{Configures}.}
\label{MRh-SLy4}
\end{figure*}
 
 \begin{figure*}
\sidecaption
\includegraphics[width=11.2cm]{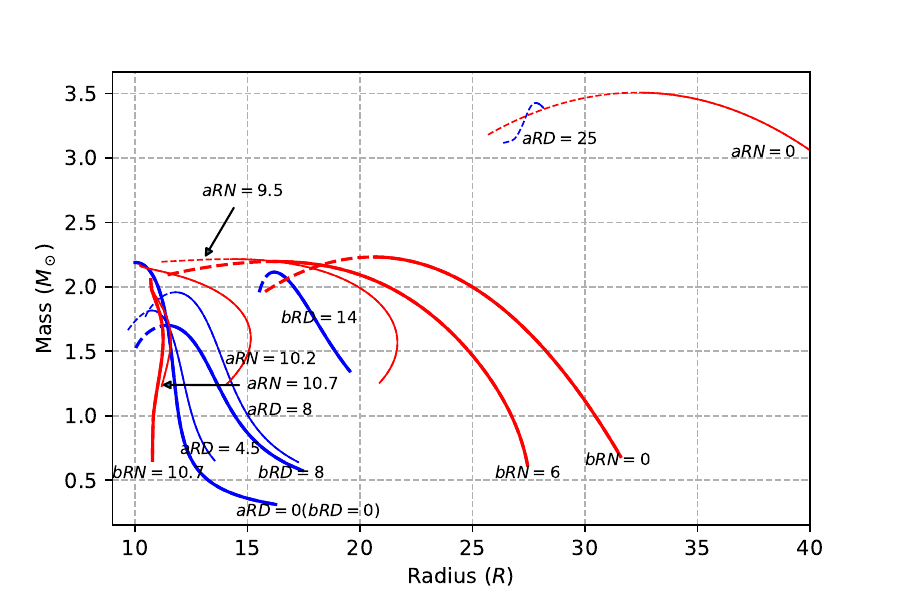}
\caption{Mass-Radius relations of the hybrid stars of type (I) and (II) in Fig. \ref{stars}, which are made of nuclear matter of APR4 EoS and dark matter of EoS \eq{DEoS} with ${\cal B}=0.035$ or $0.055$. The notations are the same with that in Fig. \ref{MRh-SLy4}.  }
\label{MRh-APR4}
\end{figure*}

\begin{figure*}
\sidecaption
\includegraphics[width=11.2cm]{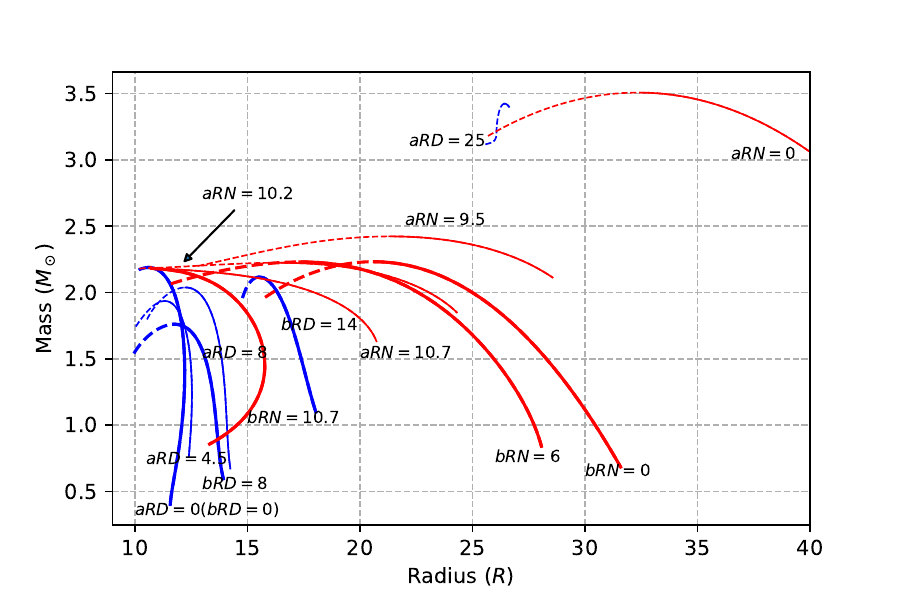}
\caption{Mass-Radius relations of the hybrid stars of type (I) and (II) in Fig. \ref{stars}, which are made of nuclear matter of SKb EoS and dark matter of EoS \eq{DEoS} with ${\cal B}=0.035$ or $0.055$. The notations are the same with that in Fig. \ref{MRh-SLy4}. }
\label{MRh-SKb}
\end{figure*}

\begin{figure*}
\sidecaption
\includegraphics[width=11cm]{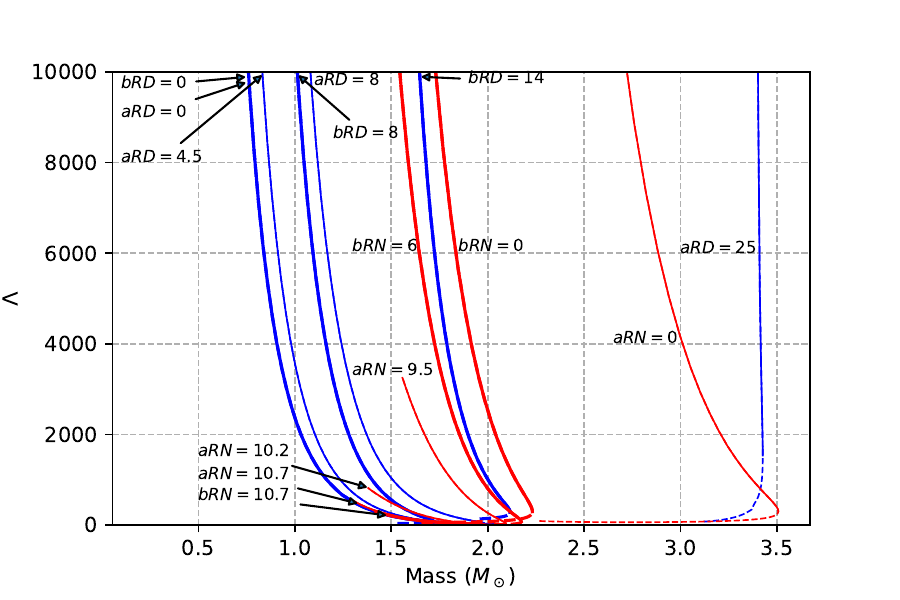}
\caption{Corresponding TLN-Mass relations of Fig. \ref{MRh-SLy4}. Similarly, the unstable configurations are indicated by the parts of dashed lines. }
\label{MLh-SLy4}
\end{figure*} 

\begin{figure*}
\sidecaption
\includegraphics[width=11cm]{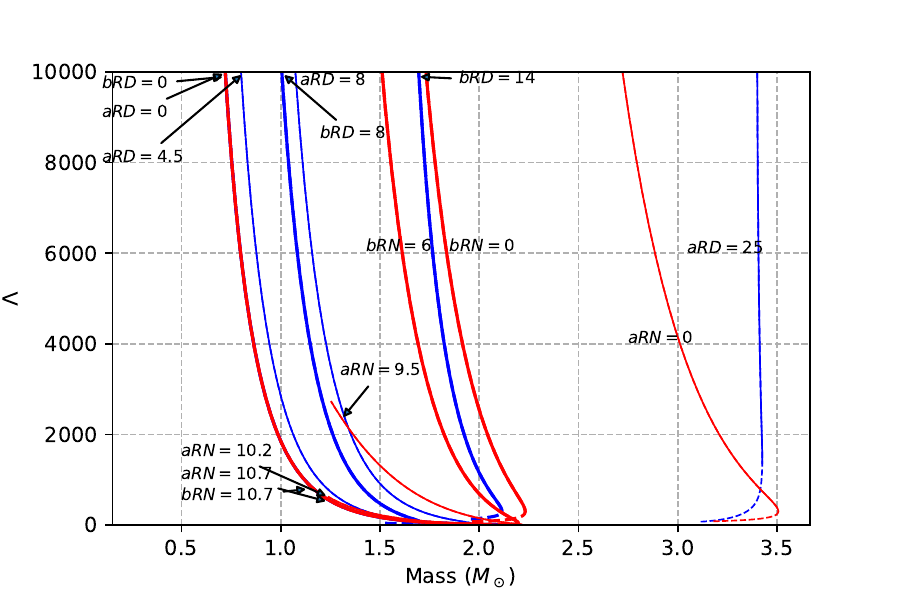}
\caption{Corresponding TLN-Mass relations of Fig. \ref{MRh-APR4}. Similarly, the unstable configurations are indicated by the parts of dashed lines. }
\label{MLh-APR4}
\end{figure*} 

\begin{figure*}
\sidecaption
\includegraphics[width=11cm]{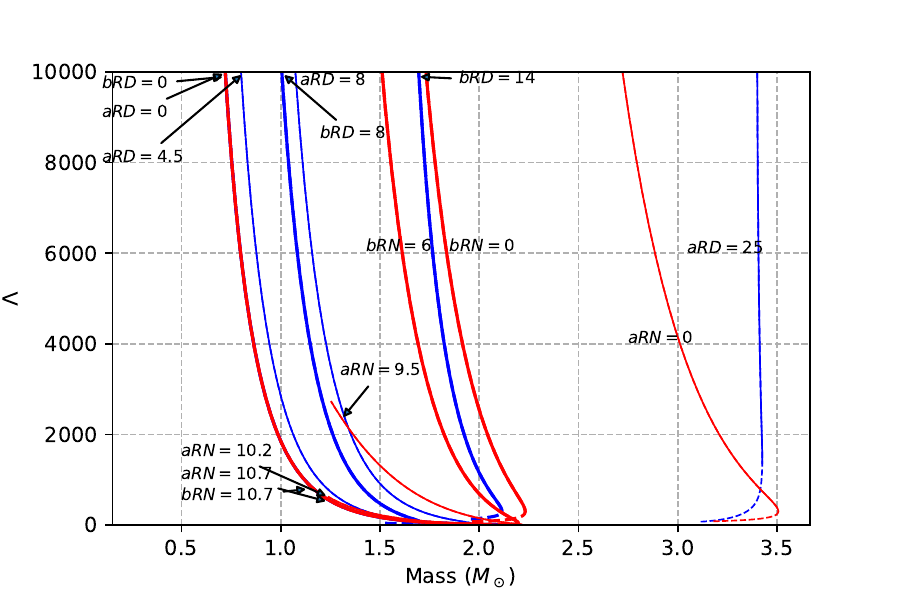}
\caption{Corresponding TLN-Mass relations of Fig. \ref{MRh-SKb}. Similarly, the unstable configurations are indicated by the parts of dashed lines. }
\label{MLh-SKb}
\end{figure*}

\begin{figure*}
\sidecaption
\includegraphics[width=12.5cm]{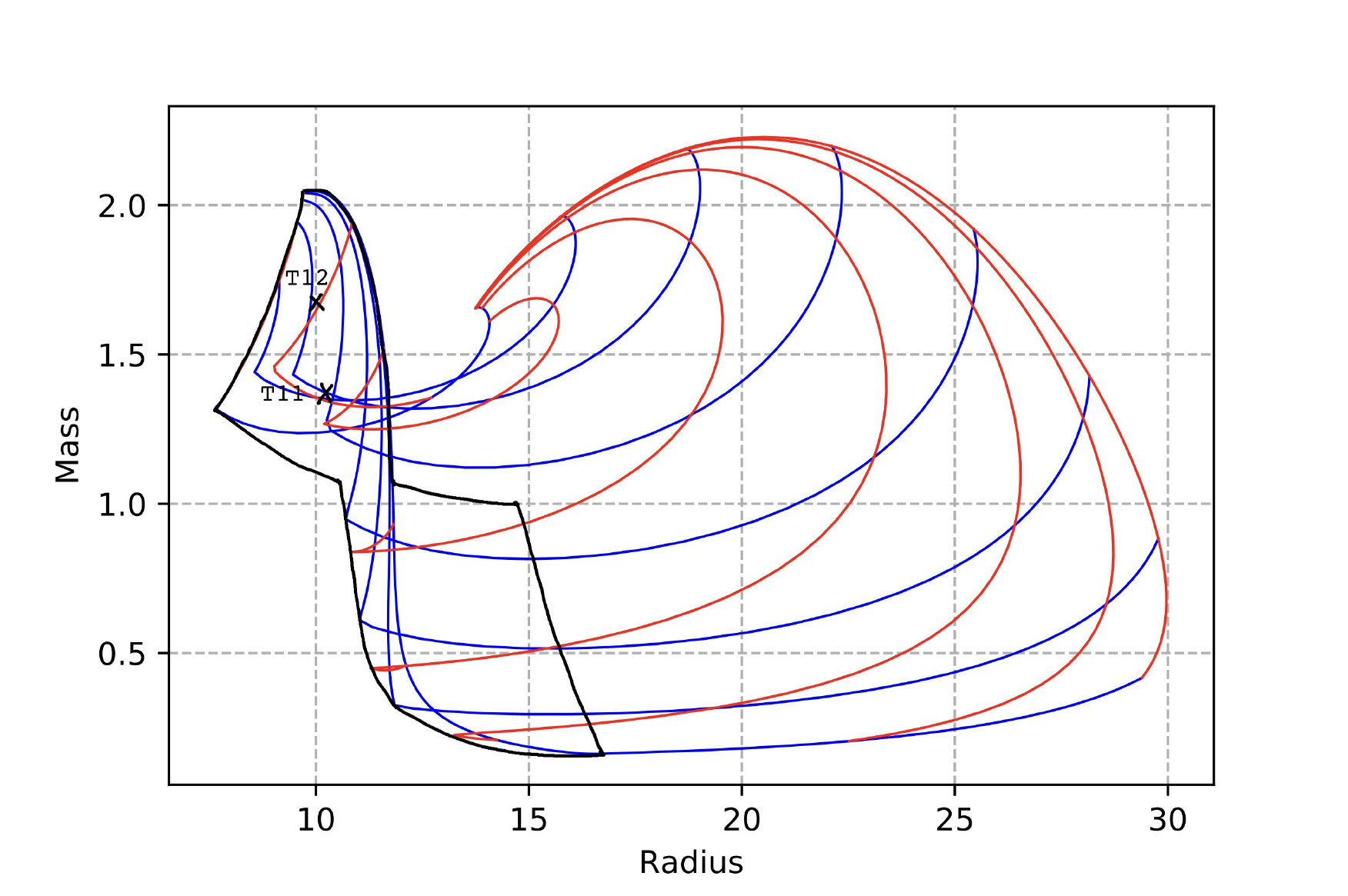}
\caption{Mass-Radius (Up)  and TLN-Mass (Down) relations of the hybrid stars of type (IIIa) and (IIIb) in Fig. \ref{stars} for SLy4 EoS and for dark matter EoS  with ${\cal B}=0.055$. The region of stable configurations is specified by the black-encircled area.  Different black lines correspond to different core pressures of dark matter but varying the core pressures of the nuclear matter, and the brown lines are the other way around. The crosses indicate special stars T11 and T12 mentioned later in Table \ref{Configures}.}
\label{Mix-SLy4}
\includegraphics[width=12.5cm]{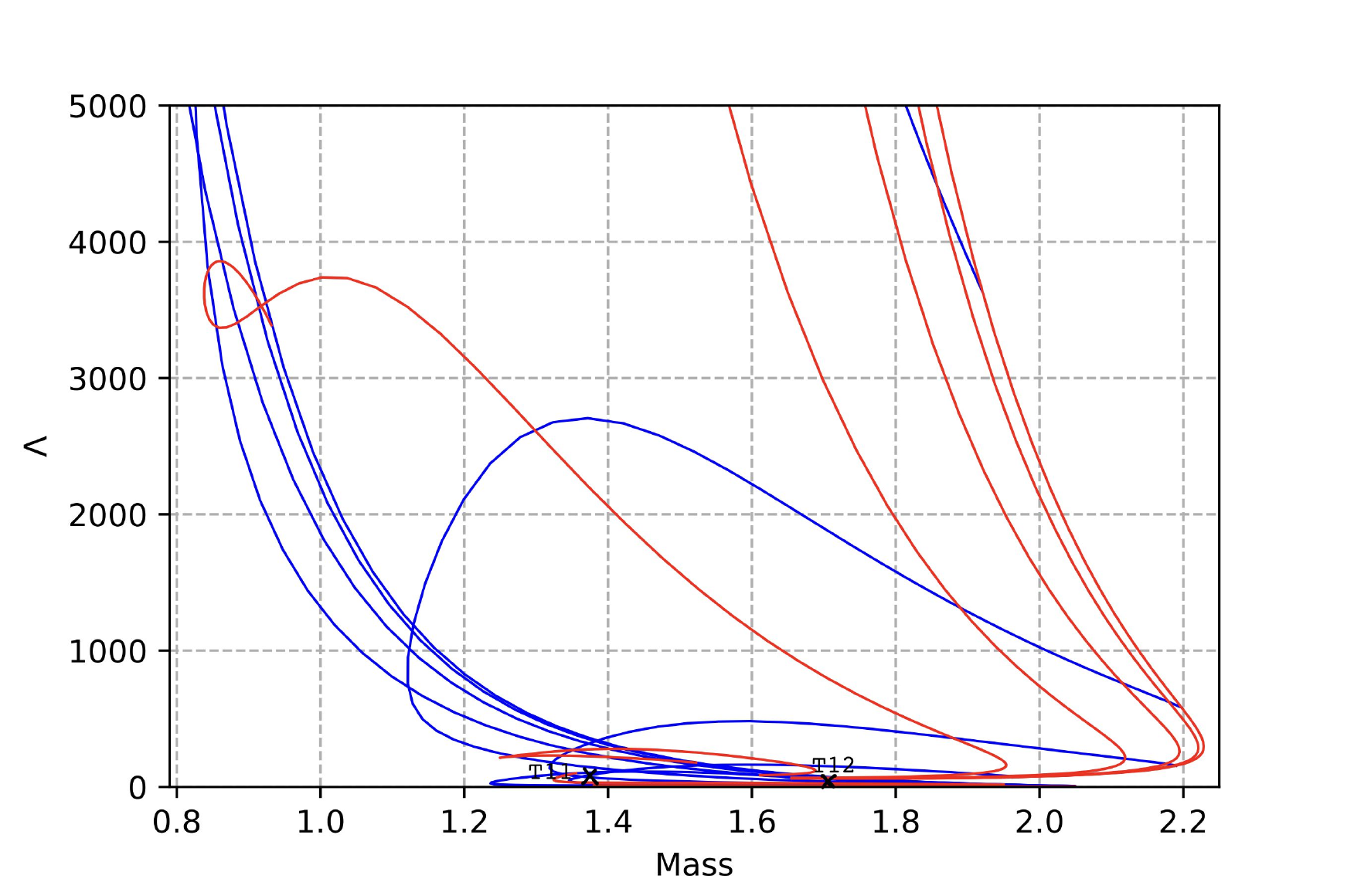}
\end{figure*}

\begin{figure}
 \centering
 \includegraphics[width=9cm]{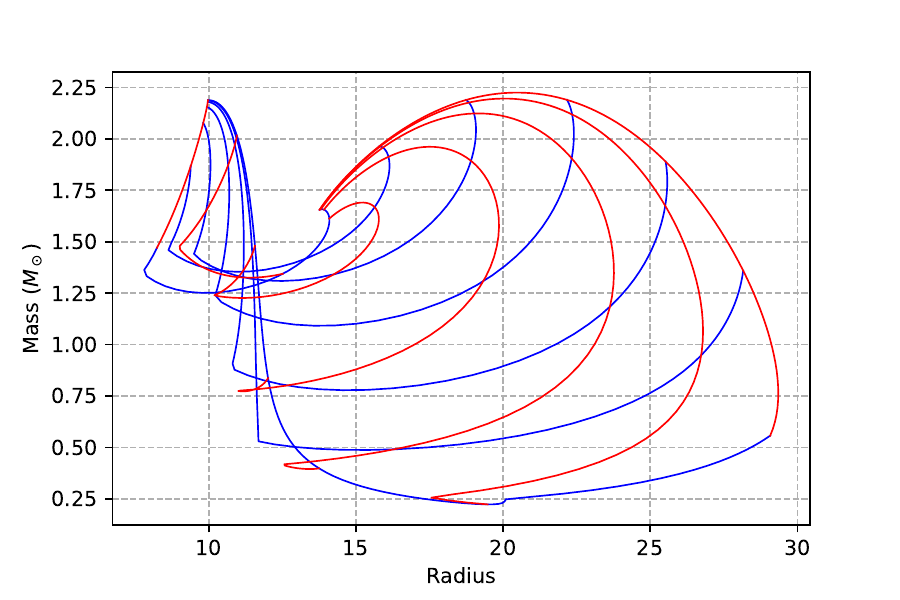}
 \hspace{1in}
 \includegraphics[width=9cm]{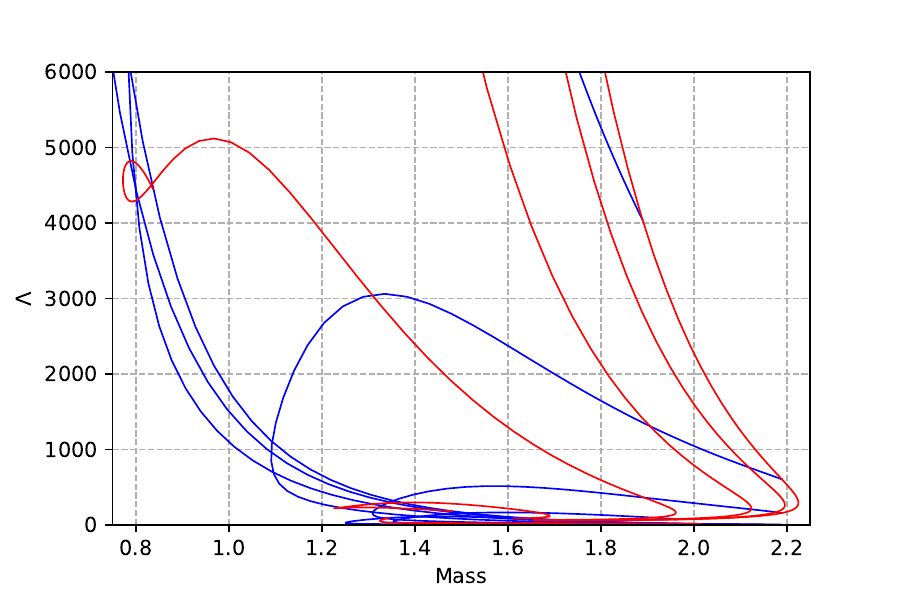}
\caption{Mass-Radius (Up)  and TLN-Mass (Down) relations of the hybrid stars of type (IIIa) and (IIIb) in Fig. \ref{stars} for APR4 EoS and for dark matter EoS  with ${\cal B}=0.055$.  The colouring means the same as that in Fig. \ref{Mix-SLy4}.}
\label{Mix-APR4}
\end{figure}

\begin{figure}
 \centering
 \includegraphics[width=9cm]{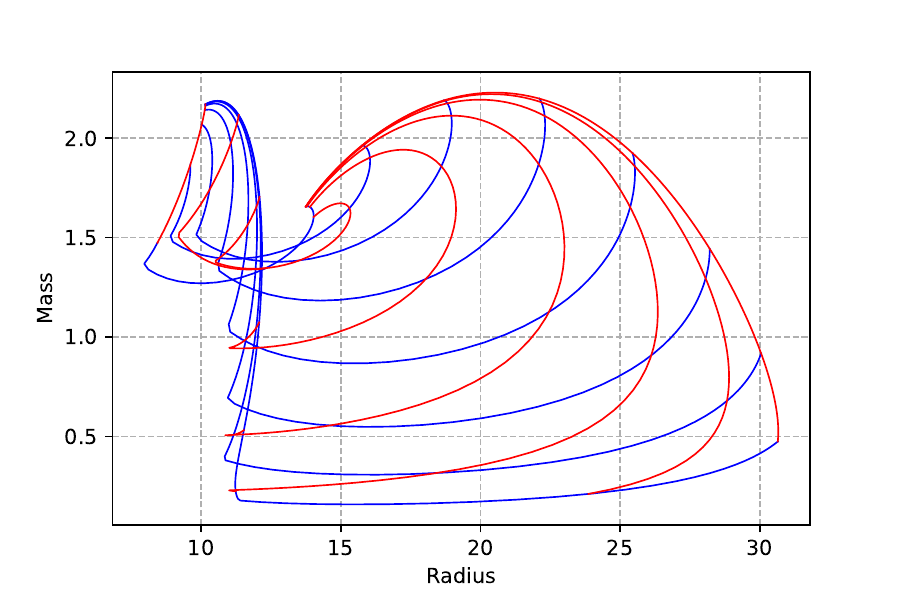}
 \hspace{1in}
 \includegraphics[width=9cm]{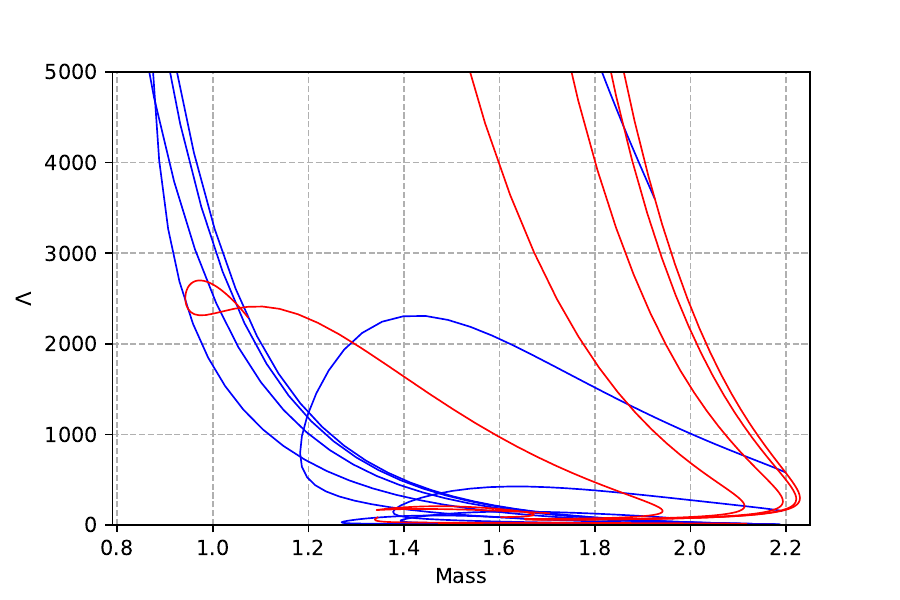}
\caption{Mass-Radius (Up)  and TLN-Mass (Down) relations of the hybrid stars of type (IIIa) and (IIIb) in Fig. \ref{stars} for SKb EoS and for dark matter EoS  with ${\cal B}=0.055$.  The colouring means the same as that in Fig. \ref{Mix-SLy4}.}
\label{Mix-SKb}
\end{figure}

  \subsection{For Scenario I and II:}\label{section 4.1}
 We consider the case with three different EoSs for nuclear matter and vary the parameters of the dark matter EoS. In Fig. \ref{MRh-SLy4} to Fig. \ref{MRh-SKb} we show the $M$-$R$ relations for the first two scenarios for ${\cal B}=0.035$ (with various $r_W$ labelled as {\bf aRN} for the first scenario, and as {\bf aRD} for the second) or ${\cal B}=0.055$ (labelled by {\bf bRN} and {\bf bRD}), and in Fig. \ref{MLh-SLy4} to Fig. \ref{MLh-SKb} we show their corresponding $\Lambda$-$R$ relations. From the results, we observe the following. 
 (i) The one labelled by {\bf aRD}$=0$ (and {\bf bRD}$=0$) is the pure neutron stars which can both fit GW170817 and the  traditional astronomical observations. As the dark matter core increases its size,  the maximal mass goes down first as the boson EoS is less stiff, then goes up as the allowed maximal mass of boson EoS is higher, and at the same time the total radius becomes larger. (ii) For the first scenario, we see that there is a jump around {\bf aRN}$=9.5$ in Fig. \ref{MRh-SLy4} and Fig. \ref{MRh-APR4}, {\bf aRN}$=10.2$ in Fig. \ref{MRh-SKb}, or {\bf bRN}$=7$ in Fig. \ref{MRh-SLy4}  (though {\bf bRN}$=7$ is not shown explicitly) respectively, beyond which the small-radius configurations become unstable (indicated by dash line in Fig. \ref{MRh-SLy4} to Fig. \ref{MRh-SKb}), this may imply some first order phase transition. On the other hand, above this critical {\bf RN}, there are more compact hybrid stars which can be consistent with LIGO/Virgo observations with small TLN as indicated in Fig. \ref{MLh-SLy4} to Fig. \ref{MLh-SKb}. (iii) The configurations with $M$ larger than $3M_{\odot}$ are mainly composed of dark matter, as seen from the ratio of $r_W/R$. Contrarily, the more compact hybrid stars of smaller $R$ are mainly composed of nuclear matter. This is understandable since the  three neutron EoSs are stiffer than EoS \eq{DEoS}. It then implies that the final states of most binary hybrid stars' mergers are unstable unless the initial stars are almost pure dark stars. Thus, if the component stars of GW190425 are the hybrid stars of these two scenarios, the final state will collapse into a black hole. 

In a short summary: in general, due to the additional component of matters, our hybrid stars can host a wider range of masses and TLNs than pure neutron stars or dark stars. This will then be taken as a special feature to distinguish from the pure neutron and dark stars of a given EoS in the forthcoming GW observational data. 
    
Finally, we should remind the readers that in scenario I and II, $r_W$ should be determined by a given model interaction between dark matter and baryons. Therefore, our results can be thought as tabulating the possible configurations of scenarios I and II for generic model dark matter-baryon interactions.

 \subsection{For Scenario III:}   
Next, we show the $M$-$R$ and $\Lambda$-$M$ relations for the third scenario of hybrid stars, namely the mixed ones in the core, and the associated star configurations are denoted by {\bf bMX} for  ${\cal B}=0.055$ of dark matter EoS combined with the three neutron EoSs. In Fig. \ref{Mix-SLy4} to Fig. \ref{Mix-SKb} we show the results.

Unlike the first two scenarios,  there are saddle instabilities for the third scenario. 
{{For a single-component star, its stability is determined by varying the central pressure. For a hybrid star of scenario III, the central pressure is a sum of the neutron partial pressure and dark matter partial pressure. One necessary condition for a stable hybrid star is that, it remains stable when perturbing either partial pressure while keeping the other fixed. This is called saddle (in)stability.}}
To judge the stable regions, we apply
 the lessons learned from {the (reverse) BTM Stability Criteria in the appendix. } According to the typical examples in Fig. \ref{mr_stablility}, we can roughly \footnote{{
By “roughly” we mean that the area boarder is not marked in high accuracy, but the stability of each point can be precisely determined if we draw the lines more densely.}}
mark the stable regimes in  Fig. \ref{Mix-SLy4} by the black-encircled areas. 
We omit to mark the stable regimes in Fig. \ref{Mix-APR4} and Fig. \ref{Mix-SKb}, since  those are similar to that in  Fig. \ref{Mix-SLy4}.

We see that the stable hybrid stars of the third scenario are limited to the left part of the $M$-$R$ curves, which are more NS-like and can have the maximum masses comparable with the ones of pure neutron stars. On the right-top part of the curves, there are no stable DM-like hybrid stars due to the saddle instability. 
 One naive interpretation is because that a boson star with low central pressure like $10^{-6} p_{\odot}$ is stable, while a neutron star with the same central pressure is unstable. Thus, a hybrid star deviating a little from a pure neutron star can have a stable boson part with low boson central pressure.
Moreover, as in the first two scenarios, our hybrid stars can feature a wider range of masses and TLNs than the pure neutron stars of a given EoS. 
   
 Even there are still uncertainties for the EoS of nuclear matter which may hinder our above interpretations, it is still interesting to demonstrate the capability of our hybrid stars with a given nuclear matter EoS, i.e., SLy4, to interpret the observed GW events such as GW190425. The full picture by pinning down the uncertainties for both nuclear and dark matter should be expected from the PE of the coming GW events.

\section{Fitting of GW170817 and GW 190425}
After discussing the general $M$-$R$ and $\Lambda$-$M$ relations for the three scenarios of compact hybrid stars, we see the three representative neutron EoS lead to similar results, so we mainly choose SLy4 case as an example for the discussion in this section.
We now pick up some specific configurations as listed in Table \ref{Configures} which can be identified as the component stars for the GW170817 and GW 190425. In Table  \ref{Configures} we have listed 12 hybrid stars labeled by the index Tn with n$=1\cdots 12$. Most of them are indicated on Fig. \ref{MRh-SLy4}  and \ref{Mix-SLy4}.   The types {\bf aRD}, {\bf aRN}, {\bf bRD}, {\bf bRN} and {\bf bMX} are defined as before to indicate the different choices of core radius and  ${\cal B}$. Especially, we list T6 and T8 to show that the typical high-mass stars with masses larger than $3M_{\odot}$ are mainly dark stars and cannot be the final states of the mergers of low mass hybrid stars mainly composed of nuclear matter. Besides, most of the stars with masses lower than 2$M_{\odot}$ have radii just 2 or 3km larger than the typical radii of neutron stars, say around 11 km. Some of them such as T7 and T10 yet have 10.7 km neutron cores to be consistent with the observed electromagnetically visible radius \citep{Lattimer:2013hma, Cackett:2007pi, Ozel:2015fia}.

\begin{table}[!htbp]
\centering
{\scriptsize
\begin{tabular}{|c|c|c|c|c|c|c|c|}
\hline
Index&Type&$M$&$M_{D}/M$&$R$&$R_{D}$&$R_{N}$&$\Lambda$\\ 
\hline
T1 & {\bf aRD} &1.37&0&11.59&0 & &360\\
\hline
T2 & {\bf aRD} &1.37&0.07&11.96&4.5& &486\\
\hline
T3 & {\bf aRD} &1.9&0.65&10.77&8& &21\\
\hline
T4 & {\bf aRD} &1.7&0.26&13.19&8& &352\\
\hline
T5 & {\bf aRD} &1.5&0.21&13.54&8& &999\\
\hline
T6 & {\bf aRD} &3.42&0.93&26.91&25& &1802\\
\hline
T7 & {\bf aRN}  &1.7&0.15&12.90& &10.7&79\\
\hline
T8  & {\bf aRN}  &3.5&1.0&33.34& &0&390\\
\hline
T9 & {\bf bRD}  &1.7&0.53&11.63&8& & 104\\
\hline
T10 & {\bf bRN}  &1.7&0.01&11.97& &10.7&77\\
\hline
T11 & {\bf bMX} &1.37&0.30&10.14&10.14& 9.35&87\\
\hline
T12 & {\bf bMX} &1.7&0.11&10.06& 7.62&10.06&25\\
\hline
\end{tabular}
}
\caption{List of 12 specific hybrid stars, most of which are indicated on Fig.  \ref{MRh-SLy4} and \ref{Mix-SLy4}.  The first entry labels the stars, and the second entry is the type of hybrid stars as defined earlier, then the subsequent entries are total mass, mass ratio of dark matter to the total mass, total radius, respective core radius and TLN. The core pressures of dark matter for  T11 and T12 are $2.5\times10^{-4} p_{\odot}$ and $2.0\times10^{-4} p_{\odot}$, respectively. } \label{Configures}
\end{table}   
   
Since $\tilde{\Lambda}=(\Lambda_1+\Lambda_2)/2$ for the equal-mass binary,  any two stars from the same type labelled by either {\bf a} or {\bf b}, e.g., two T2's, two T11's, or $\{$T1, T2$\}$ etc, can form a binary of hybrid stars with  $\tilde{\Lambda}$ close to it observational upper bound to explain GW170817.  
  
In contrast, for GW190425 the inferred total mass $M_1+M_2 \simeq 3.4^{+0.3}_{-0.1}M_{\odot}$, with $M_1\in (1.62,1.88) M_{\odot}$, $M_2\in (1.45,1.69) M_{\odot}$ and $\tilde{\Lambda}\le 600$ for low spin prior, or    with $M_1\in (1.61, 2.52) M_{\odot}$, $M_2\in (1.12, 1.68) M_{\odot}$ and $\tilde{\Lambda}\le 1100$ for high-spin prior \citep{Abbott:2020uma}. From Table \ref{Configures}, we can find the following pairs of hybrid stars with SLy4 EoS to explain GW190425:  (1)  two T4's  with $\tilde{\Lambda}=352$, (2)  $\{$T3, T5$\}$ with $\tilde{\Lambda}=348$, (3)  two T7's  with $\tilde{\Lambda}=79$, (4)  two T9's  with $\tilde{\Lambda}=104$, (5)  two T10's  with $\tilde{\Lambda}=77$ and  (6)  two T12's  with $\tilde{\Lambda}=25$. From Table  \ref{Configures} it is interesting to see that the values of ${\Lambda}$ cover a wide range, even with the same masses.

Note that for GW170817, the inferred total mass $M_1+M_2 \simeq 2.73^{+0.04}_{-0.01}M_{\odot}$ with $M_1\in (1.36,1.60) M_{\odot}$, $M_2\in (1.16,1.36) M_{\odot}$ and $\tilde{\Lambda}=300^{+420}_{-230}$ for low-spin prior. For simplicity, we consider the equal mass pair with $M_1=M_2=1.37M_{\odot}$ \citep{TheLIGOScientific:2017qsa,Abbott:2018wiz}. The pure-neutron or hybrid stars with such mass in the list are T1, T2 and T11. In this set, unlike T11 which belongs to the third scenario, T1 and T2 belong to the first two scenarios and have none or little dark matter.

In conclusion, we have demonstrated that it is quite easy to fit the gravitational events GW170817 and \\GW190425 as some hybrid stars of all three scenarios considered in this paper. This opens a venue to fit the future gravitational events as the hybrid stars considered in this paper, and can further pin down the possible parameter range of $m$ and $\lambda$ of $\phi^4$ SIDM by the method introduced in the next section.

\section{Parameter estimation for EoS of dark matter}   

We finally would like to demonstrate that one can use the GW data and the associated parameter estimation (PE) results to obtain the posteriors of the EoS parameter $\cal B$ and the model parameter $r_W$ of the hybrid stars. For simplicity, we will consider the first two scenarios of hybrid stars made of nuclear matters with three kinds of EoS and the bosonic SIDM with EoS given by \eq{DEoS}. {Based on the results, we can further determine the range of the model parameters $m$ and $\lambda$ of $\phi^4$ SIDM. This shows that the GW data analysis can be used to constrain the model of dark matter, mark as a triumph of GW astroparticle physics.} 

\begin{figure}[htb]
\resizebox{\hsize}{!}{\includegraphics{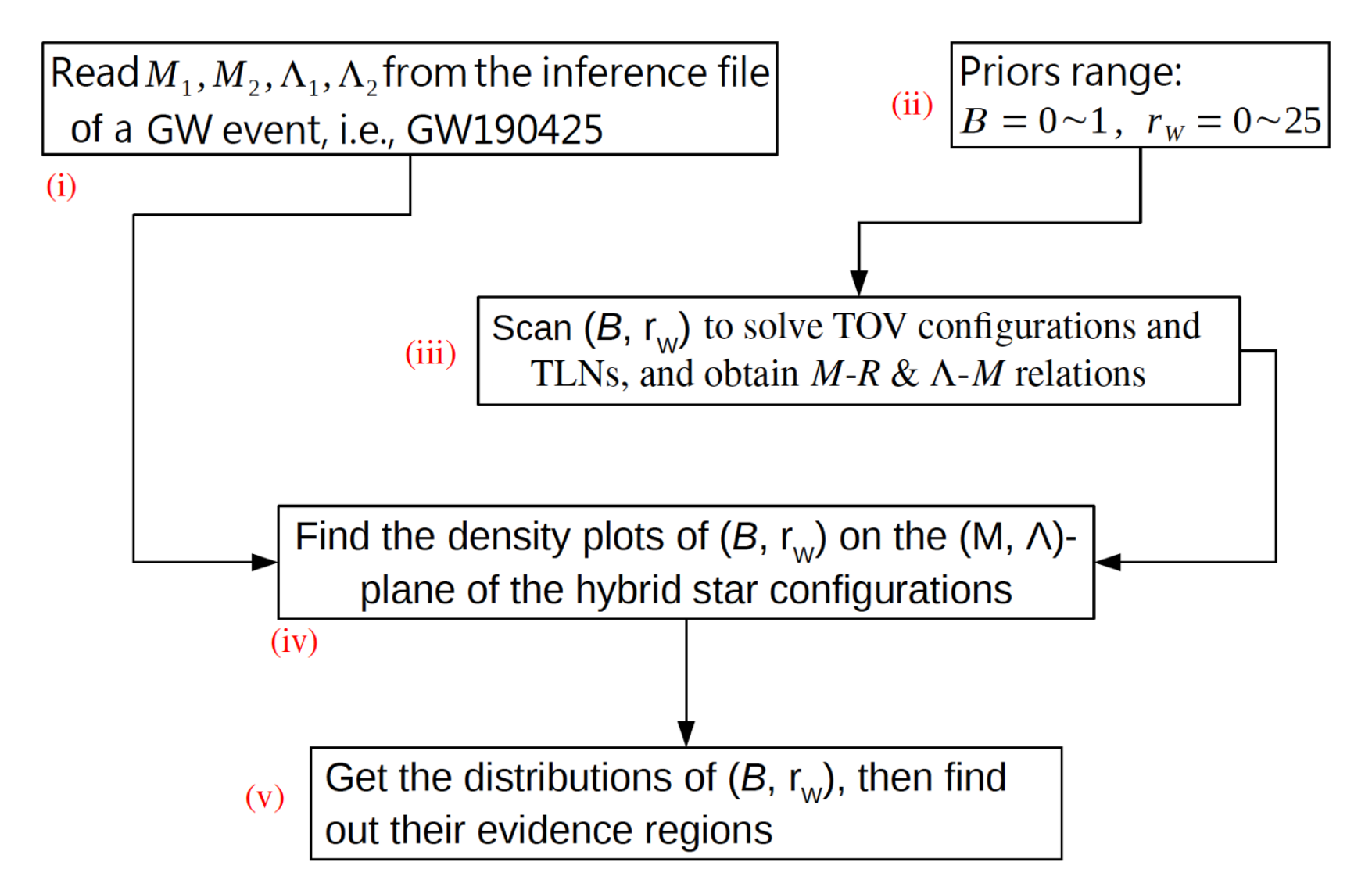}}
\caption{\label{eos-fit}
Flow chart of obtaining posteriors of model parameters for the hybrid stars from the posteriors of PE for a particular GW event, such as GW190425. {Here, Box (i) is to obtain the headcounts of the PE posterior of GW190425; Box (ii) and (iii) are to find the corresponding $({\cal B}, r_W)$ for each headcount labelled by $(M,\Lambda)$ from Mass-Radius and TLN-Mass relations; and Box (iv) and (v) are to use the results in the previous step to obtain the density plot of $({\cal B}, r_W)$ on the $(M,\Lambda)$-plane, and their corresponding evidence regions and marginal distributions. See also the discussions in the main text.} }
\end{figure}

{The main idea is to map the headcounts, which are labelled by $(M,\Lambda)$, of the PE posteriors of a GW event based on Markov-Chain-Monte-Carlo method into the density plots of $\cal B$ and $r_W$, via the help of the Mass-Radius and TLN-Mass relations obtained in section \ref{section 4}. The headcount here means the number of populations of a particular TOV configuration labeled by $(M,\Lambda)$ in the inference file associated with the PE posteriors from MCMC sampling. Given the model parameters $({\cal B},r_W)$, one can solve a set of stable hybrid stars specified by the Mass-Radius and TLN-Mass relations. Therefore, by associating the headcounts in the PE posteriors with possible $({\cal B},r_W)$'s one can then obtain the density plots of $\cal B$ and $r_W$ on the $(M,\Lambda)$-plane. In summary, the key procedures of the above mapping are summarized in Fig. \ref{eos-fit}, each Box of which performs some intermediate steps of the task. These steps go as follows: (1) Box (i) in Fig. \ref{eos-fit} is to obtain the headcounts of the PE posterior of GW190425 from \citep{Abbott:2020uma}; (2) Box (ii) is to discrete the prior space of $({\cal B}, r_W)$;  (3) Box (iii) is to solve the TOV configurations and TLN for each point in this discrete prior space; (4) using the results of (3), Box(iv) is to obtain the density plot of $({\cal B}, r_W)$ on the $(M,\Lambda)$-plane; (5) Box (v) is to obtain the associated evidence regions and marginal distributions of $({\cal B}, r_W)$.}

\begin{figure}[htb]
\resizebox{\hsize}{!}{\includegraphics{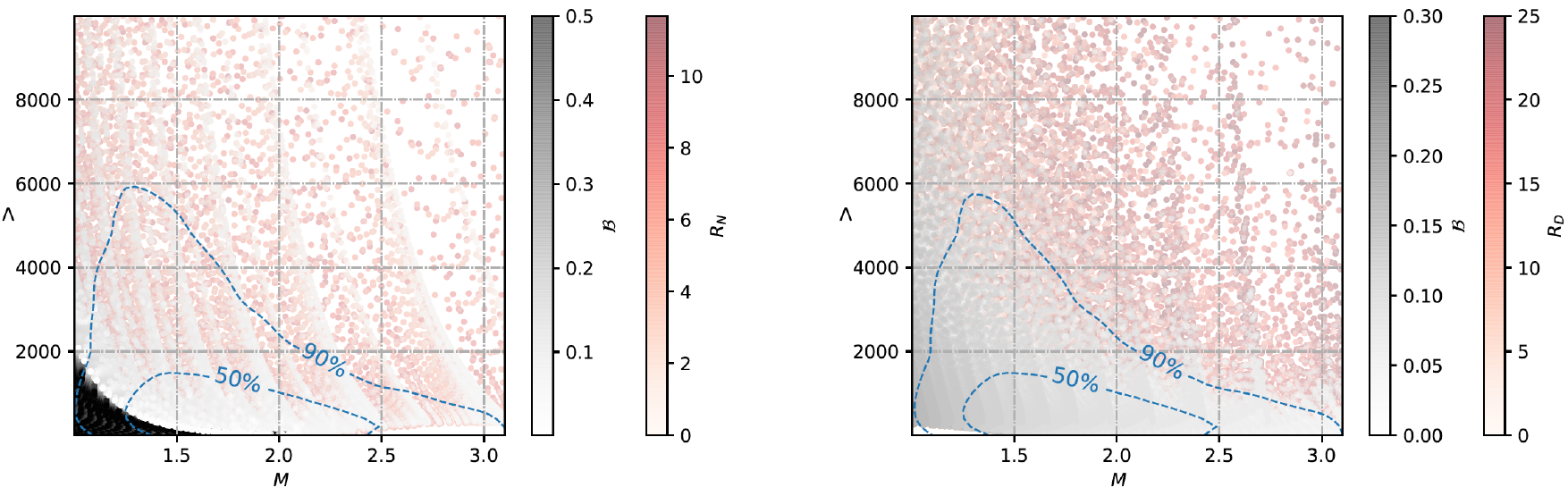}}
\caption{Density plots of parameters $({\cal B}, r_W)$ on the $(M,\Lambda)$-plane of the hybrid star configurations of scenario I (Left panel, with $r_W=R_N$) and II (Right panel, with $r_W=R_D$), with neutron EoS chosen as SLy4. The sidebars indicate the scales of the density plots. There is no hybrid star configuration falling inside the corresponding $(M,\Lambda)$ regions when $({\cal B}, r_W)$ go beyond the maximal values of the sidebars. The credible intervals of 50\% and 90\% of the PE results for GW190425 are indicated by the blue dashed lines  \citep{Vallisneri:2014vxa}. 
} \label{gw190425-m-l}
\end{figure}

\begin{figure}%
{\resizebox{\hsize}{!}{\includegraphics{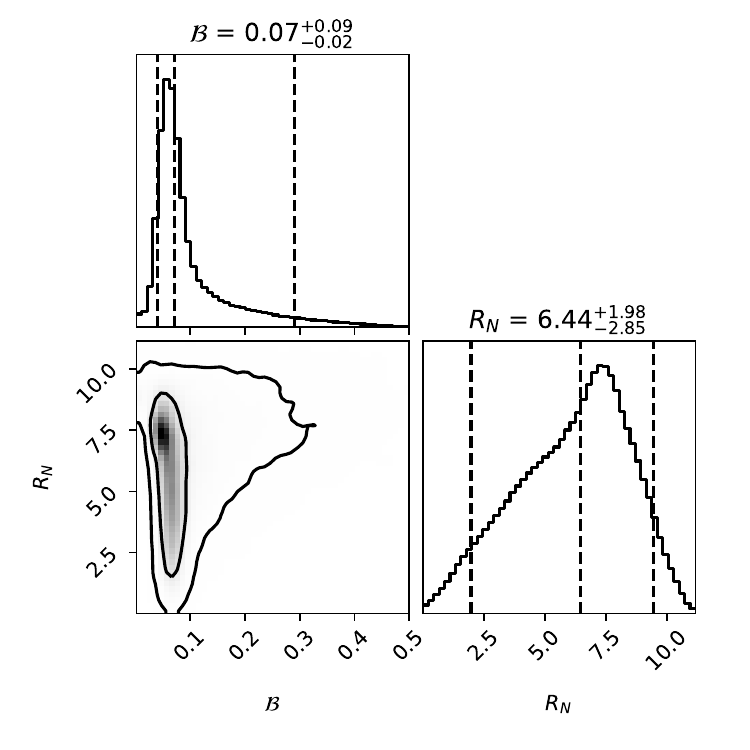}} }%
    \quad
{\resizebox{\hsize}{!}{\includegraphics{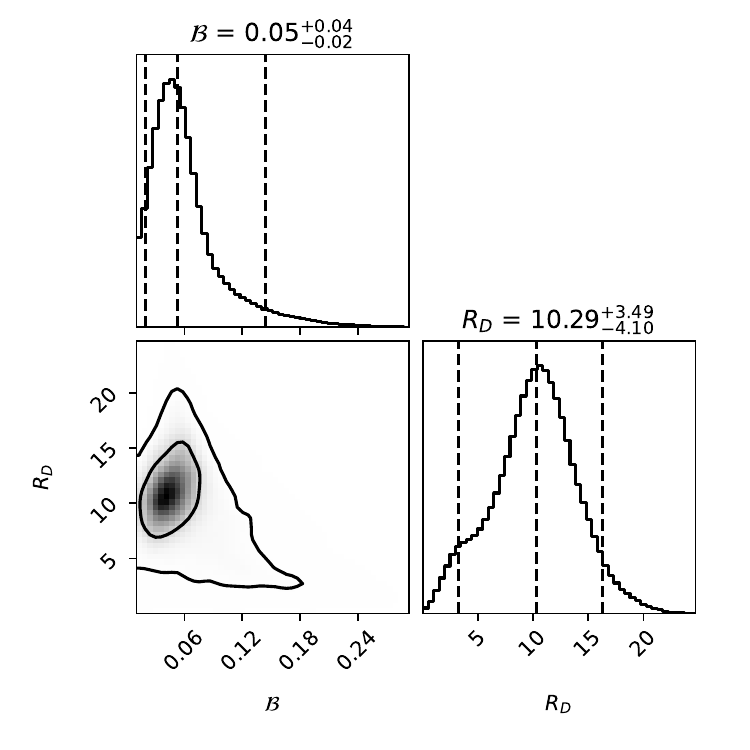}} }%
\caption{Posteriors for the dark matter EoS parameter $\cal B$ (horizontal axis) of the SIDM and the core-radius $r_W$ (vertical axis) for the first (top sub-figure) and the second (bottom sub-figure) scenarios with neutron SLy4 EoS. The inner circled regions are $50\%$ credible interval, and the outer ones are $90\%$ credible interval. The inferred best-fitted values of $\cal B$ and $r_W$ are also given in Table \ref{PE}. }%
    \label{PEforB}%
\end{figure}

\begin{figure}%
{\resizebox{\hsize}{!}{\includegraphics{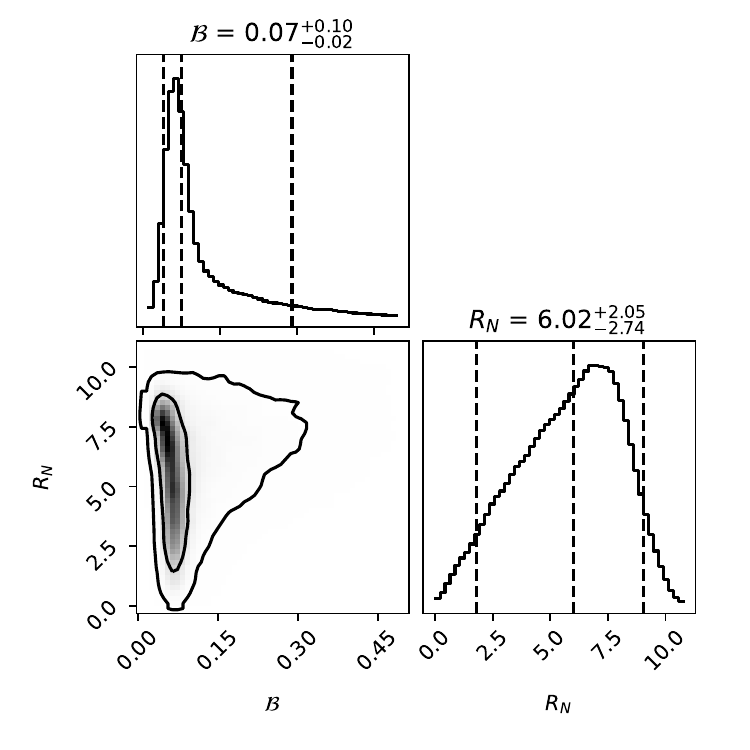}} }%
    \quad
{\resizebox{\hsize}{!}{\includegraphics{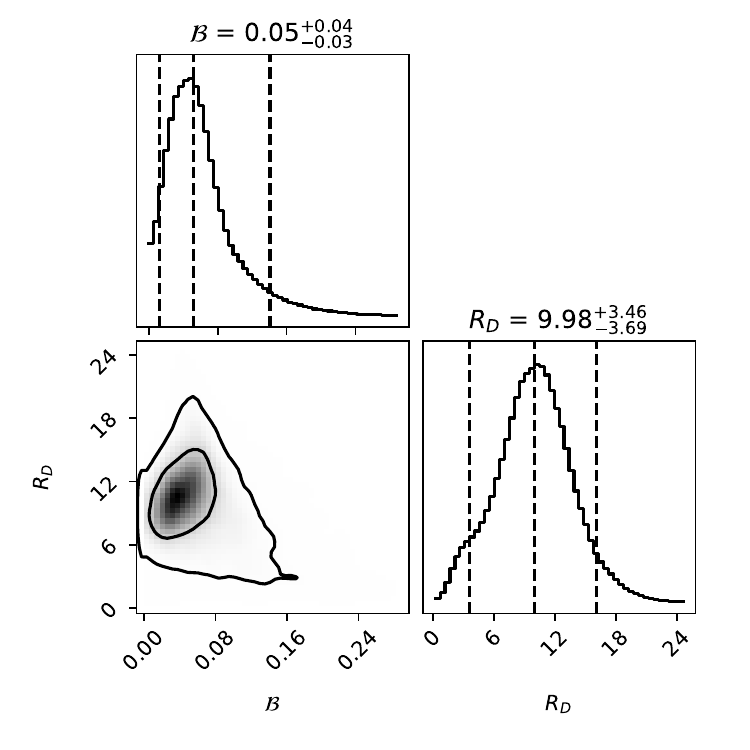}} }%
\caption{Posteriors for the dark matter EoS parameter $\cal B$ (horizontal axis) of the SIDM and the core-radius $r_W$ (vertical axis) for the first (top sub-figure) and the second (bottom sub-figure) scenarios with neutron APR4 EoS. The inferred best-fitted values of $\cal B$ and $r_W$ are also given in Table \ref{PE}. }%
    \label{PEforBAPR}%
\end{figure}

\begin{figure}%
{\resizebox{\hsize}{!}{\includegraphics{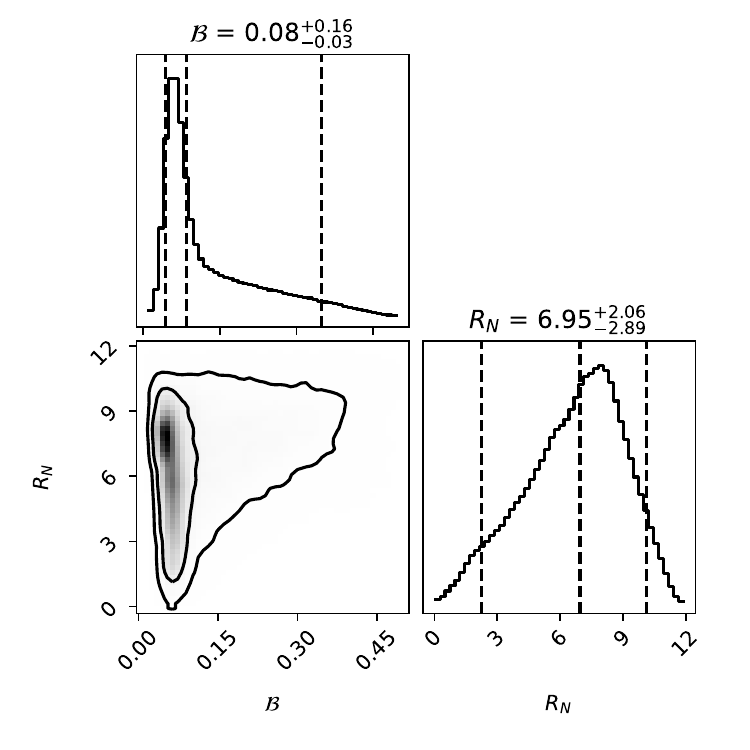}} }%
    \quad
{\resizebox{\hsize}{!}{\includegraphics{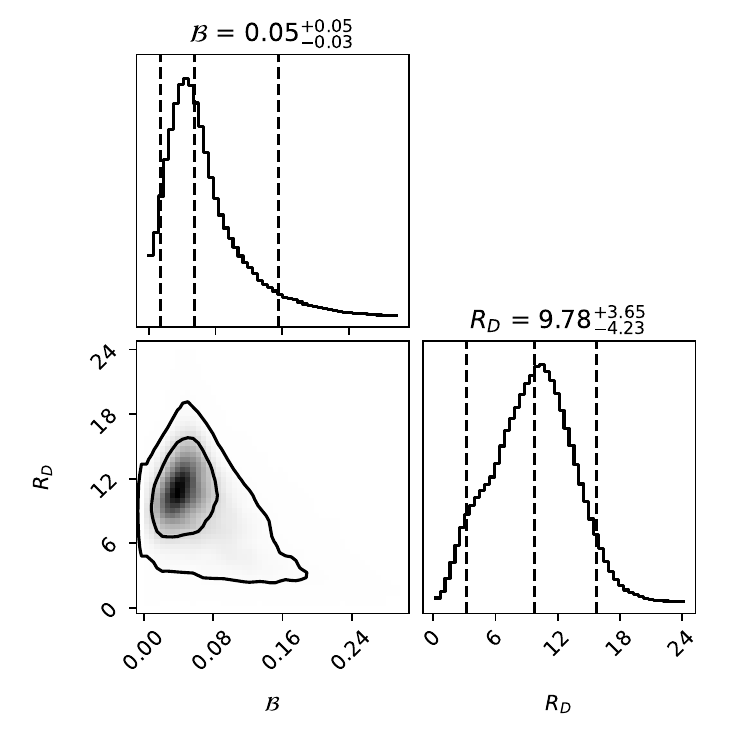}} }%
\caption{Posteriors for the dark matter EoS parameter $\cal B$ (horizontal axis) of the SIDM and the core-radius $r_W$ (vertical axis) for the first (top sub-figure) and the second (bottom sub-figure) scenarios with neutron SKb EoS. The inferred best-fitted values of $\cal B$ and $r_W$ are also given in Table \ref{PE}. }%
    \label{PEforBSKb}%
\end{figure}

{The details of the Mass-Radius and TLN-Mass relations are already obtained in section \ref{section 4}, which will then be used in step (2) of the above procedure. As mentioned, we will only consider the hybrid stars of scenarios I ($r_W=R_N$) and II ($r_W=R_D$) for estimating the $m$ and $\lambda$ from the inference file of GW190425. Follow the above procedure and the flow chart of Fig. \ref{eos-fit} by giving the prior range of  $\mathcal{B} =[0,1]$ and $r_W = [0,25]$, we obtain the density plots of $({\cal B}, r_W)$ on the $(M,\Lambda)$-plane of the hybrid star configurations of scenario I ($r_W=R_N$) and II ($r_W=R_D$) as shown in Fig. \ref{gw190425-m-l}, for the SLy4 case (other cases are similar). Based on the density plot we can further obtain the corresponding evidence region and the marginal distributions of $\cal B$ and $r_W$, with different choices of neutron EoS: SLy4, APR4 and SKb. The results are shown in Fig. \ref{PEforB} to Fig. \ref{PEforBSKb}, and  the inferred best-fitted values are summarized in Table \ref{PE}.

\begin{table}[!htbp]
\centering
\begin{tabular}{|c|c|c|c|c|}
\hline
TYPE& \multicolumn{2}{c|}{Scenario I}& \multicolumn{2}{c|}{Scenario II} \\
\hline
& ${\cal B}$ &${\bf RN} ( \mbox{km})$&${\cal B}$&${\bf RD}( \mbox{km})$\\
\hline
SLy4& $0.07^{+0.09}_{-0.02}$ &$6.44^{+1.98}_{-2.85}$&$0.05^{+0.04}_{-0.02}$&$10.29^{+3.49}_{-4.10}$\\
\hline
APR4& $0.07^{+0.10}_{-0.02}$ &$6.02^{+2.05}_{-2.74}$&$0.05^{+0.04}_{-0.03}$&$9.98^{+3.46}_{-3.69} $\\
\hline
SKb& $0.08^{+0.16}_{-0.03}$ &$6.95^{+2.06}_{-2.89}$&$0.05^{+0.05}_{-0.03}$&$9.78^{+3.65}_{-4.23}$\\
\hline
\end{tabular}
\caption{The inferred best-fitted values of the EoS parameter ${\cal B}$ and the core-radius $r_W$ for scenario I \& II of hybrid stars, with three different choices of neutron EoS. } \label{PE}
\end{table}   
We see that the PE posteriors of GW190425 give a range of the core radius of the hybrid stars of scenarios I and II. Overall, we see that the neutron core is smaller than the dark core. Moreover, note that $\bf RN$ denotes the electromagnetically visible size of the neutron core, and $\bf RD$ the size of the invisible dark core. }

{Moreover, the model parameter ${\cal B}\sim {0.08 \over \sqrt{\lambda}}({m\over \textrm{GeV}})^2$, together with the astrophysical constraint \eq{SIDMw}, i.e., $30 ({m\over \textrm{GeV}})^{3/2}$ $<$ $\lambda < 90 ({m\over \textrm{GeV}})^{3/2}$,  the above evidence region of $\cal B$ (since the results for the three neutron EoS are consistent, we take the SLy4 case for example) can yield the following compatible range of $m$ and $\lambda$ of $\phi^4$ SIDM:
\be
2.68< m < 10.53 \mbox{GeV}, \qquad 131< \lambda <3076
\ee
for scenario I, and 
\be
1.78 < m < 6.65 \mbox{GeV}, \qquad 71< \lambda <1542
\ee
for scenario II. Note that both cases satisfy well the constraint $\lambda M^2_{planck} /m^2 \gg 1$. We expect future GW data will sharpen the above estimation.  The above procedure can be also applied to other types of dark matter models, and can provide the constraints from GW data analysis for the model-building of the dark matter. Thus, we have demonstrated the possibility of constraining the dark matter model by the hybrid star scenarios. }

\section{Conclusion}   
 
In the current cosmology scenario, dark matter is more abundant than the normal matter, despite that we have known almost nothing about it. Although the self-interaction of dark matter could be weak, we cannot exclude the possibility for it to form compact objects like dark stars. The lack of traditional astronomical observations of dark stars, on the contrary, indicates the importance of GW events as probes for them. Therefore, we examine the GW properties of dark or hybrid stars by assuming their existence, rather than ruling them out for granted. We choose some special EoSs for neutrons and dark bosons in this paper, but the method we develop here can be applied to any EoS. That means, any compact star with two or more components can be discussed using our model.

In this paper we have shown that it is possible to explain some GW events such as GW170817 and GW190425 by our hybrid star scenarios. There remain uncertainties in the EoS of nuclear matter, which may hinder the hybrid star identifications. However, for a given EoS of nuclear matter, the possible configurations on the mass-radius and mass-TLN diagrams are broader than the given pure neutron star configurations. This should be true even taking into account the uncertainties of EoS for nuclear matter. Besides, the introduction of three different neutron EoSs helps to reduce the effect of the uncertainties. Therefore, we expect the dark/hybrid star scenarios should be pinned down or ruled out by future GW observations, especially when the interpretation of exotic compact objects is invoked. 
 
 

Our study integrates three areas: gravitational waves, astronomy, and particle physics. The vision becomes more clear than confined to a single subject. From our results, we see that a hybrid star can have a mass higher than the maximal mass of the neutron stars for a given EoS of nuclear matter. Moreover, the visible neutron core of this heavier hybrid star could still be comparable to the radius of the corresponding neutron star even if its total size is far larger than its core size. This then opens a possibility to lower/relax the maximal mass requirement for a nuclear EoS, and at the same time is able to interpret  the 
more massive compact object as a hybrid star when its mass is higher than the maximal mass of a conventional neutron star. A recent example is the companion star in GW190814, which has a mass of 2.6$M_{\odot}$. This is larger than the conventional neutron star's maximal mass, which is about 2.5$M_{\odot}$. Our hybrid star scenario can easily explain GW190814 as well as the other two possible BNS/BHS events, GW170817 and GW190425.


\begin{acknowledgements}
FLL, GZH, and JST are supported by Taiwan Ministry of Science and Technology (MoST) through Grant No.~106-2112-M-003-004-MY3. KZ (Hong Zhang) is supported by MoST through Grant No.~107-2811-M-003-511. We thank Yen-Hsun Lin, Alessandro Parisi and other TGWG members for helpful discussions. We also thank NCTS for partial financial support. 
\end{acknowledgements}         

\begin{appendix}
{\section{ BTM Stability Criteria}} \label{App-A}
Here we like to elaborate on the criterion for judging the stable regions of stars. A simple way of determining the stable region is the so-called BTM (Bardeen-Thorne-Meltzer)  criteria\footnote{This was shown in \citep{1966ApJ...145..505B} to be equivalent to the stability analysis by solving the Sturm-Liouville eigenmodes of radial oscillation.} \citep{1966ApJ...145..505B} as follows: We start with the stable configuration with very low core pressure, and trace along the $M$-$R$ curve by increasing the core pressure. Then, when passing through an extremum on the $M$-$R$ curve, we can have the following two situations: (1)  If the $M$-$R$ curve bends counterclockwise, a stable mode will become unstable; (2) Otherwise, one unstable mode becomes stable. By this way, we can determine which part on the $M$-$R$ curve admits stable configurations.    
    
\begin{figure}[htpb]
\resizebox{\hsize}{!}{\includegraphics{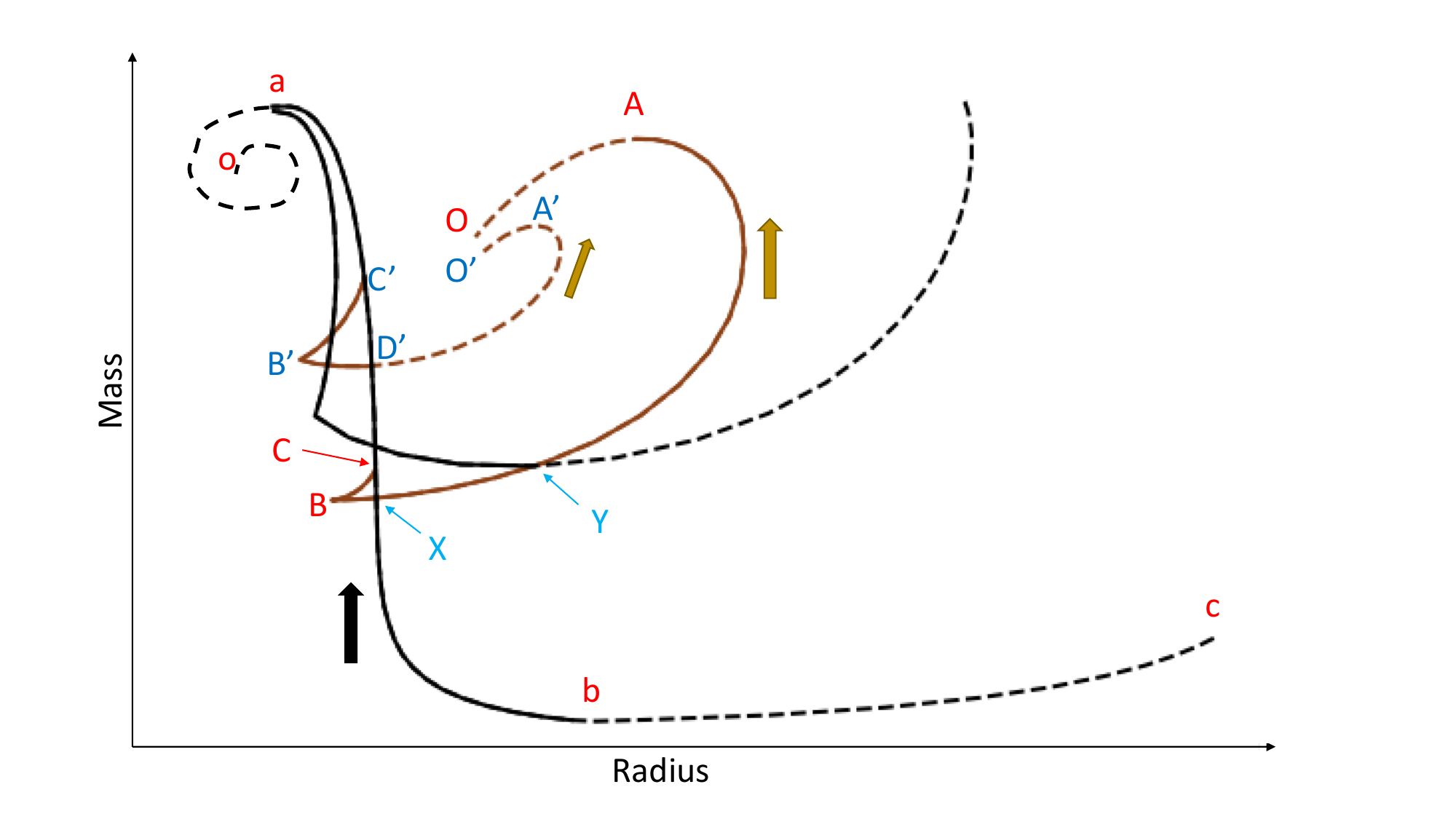}}
  \caption{Some typical examples for the stability regimes of the $M$-$R$ curve of the third scenario hybrid stars.
  The black curves are NS-like and the brown curves are DM-like, which are defined in the main text. The block arrows indicate the directions of increasing of the core pressures. We use the (reverse) BTM criteria discussed in the main text to determine the stable/unstable regimes, which  are indicated by solid/dotted parts. A caveat for the intersection point is as follows. Since the mass and radius depend on both the core pressures of nuclear and dark matter, some intersection points may be the faked ones and cannot be used to judge the saddle (in)stability. For examples, the intersection points C and Y are real ones, while the point X is a fake intersection point. For more details, see the main text.}\label{mr_stablility}
\end{figure}

However, in most of the cases we will not solve the $M$-$R$ curve for the very low core pressure, such as the case considered here. To determine the stability of the regime interested, we can assume the stability/instability of a certain part of the $M$-$R$ curve, and then apply the BTM criteria reversely as follows,
\\
{\bf Reverse BTM Stability Criteria}\footnote{{See \citep{Zhang:2020dfi} for more discussions.}}: when passing through each extremum of the $M$-$R$ curve in the direction of decreasing the core pressure,\\ 
{\bf 1.} if the $M$-$R$ curve bends clockwise, one unstable mode becomes stable;\\
{\bf 2.} Otherwise, one stable mode becomes unstable.
\\
 After that, we can apply the BTM criteria on the same regime for consistency check to determine the stability/instability of the initial part.  Some examples for the above practice are shown in Fig. \ref{mr_stablility} where the solid lines denote the stable regions, and the dotted parts the unstable ones. The arrows indicate the direction of increasing core pressure.  Note that curves $O$-$A$-$B$-$C$ and $O'$-$A'$-$D'$-$B'$-$C'$ look quite similar but differ by the extremum $D'$. Thus, they have quite different stability/instability structures after applying the above (reverse) BTM criteria.

Moreover, in the third scenario of hybrid stars, we have two orthogonal ways of changing the core pressures, and thus arrive at two sets of $M$-$R$ curves. One set called NS-like is to fix the core pressure of dark matter, but change the one of the nuclear matter, and the other one, called DM-like, is the other way around. A typical example is shown in Fig. \ref{mr_stablility}, where the black curves are NS-like and the brown curves are DM-like.  We apply the above (reverse) BTM criteria to determine the stability/instability regime of each curve, then look for the regimes where both NS-like and DM-like curves admit stability. These regimes will then be identified as the stable hybrid stars of scenario III. However, there is one caveat. In Fig. \ref{mr_stablility} we see that one brown curve may intersect one black curve twice, for example, $C$ and $X$ on the curve $O$-$A$-$B$-$C$. Since on the same black curve, the core pressure of the dark matter part is fixed, so one of them will be the fake intersection point. In this case, $X$ is not the ``real" intersection point since $C$ is the starting point at which the core pressure of the dark matter is equal to the one on the black curve.  Thus, the intersection point $X$ will not be used to judge the saddle stability. Another subtle issue is if there is a change of stability around the sharp edge points, such as $B$ and $B'$ at which a first-order phase transition from a nuclear crust to a dark crust may happen. Notice that the rotation directions before and after such points are opposite (from clockwise to counterclockwise) thus the BTM criteria cannot be applied, and $B'$ is not even an extremum. In \citep{Alford:2017vca} it was shown that there is a stability change around such points. More rigorous derivation is needed for the future study. However, due to the fact that  $B'$-$D'$ and $B$-$A$ are already stable, $C'$-$D'$ and $C$-$B$ shall be stable no matter if we adopt the criteria of \citep{Alford:2017vca}. 

\end{appendix}


\bibliographystyle{spphys} %
\bibliography{ZHL.bib} 

\end{document}